# Cryptologic Techniques and Associated Risks in Public and Private Security.

An Italian and European Union Perspective with an Overview of the Current Legal Framework[1].

by Zana Kudriasova

**Abstract.** This article examines the evolution of cryptologic techniques and their implications for public and private security, focusing on the Italian and EU legal frameworks. It explores the roles of cryptography, steganography, and quantum technologies in countering cybersecurity threats, emphasising the need for robust legislation to address emerging challenges. Special attention is given to Italy's legislative reforms, including Law No. 90 of 2024, which strengthens penalties for cybercrimes and establishes the National Cryptography Centre within the Italian National Cybersecurity Agency. Additionally, the article highlights international initiatives, such as the UN's draft convention on cybercrime, emphasising the balance between security, privacy, and fundamental human rights in a post-quantum era.

**Acknowledgements.** I extend my deepest gratitude to the Scientific Commission of *Rivista di Polizia* and its editor, Ugo Pioletti, as well as *Giustizia Penale*, for providing me with the invaluable opportunity to delve deeply into this research. I am also profoundly grateful to Prof. Leonardo Mazza from the Sapienza Universita` di Roma and Prof. Andrei Yafaev from the University College London for their guidance, encouragement, and insightful feedback throughout this research.

---

[1] This is the English version of the article, titled *Le tecniche crittologiche ed i rischi per la sicurezza pubblica e privata* was published in *La Giustizia Penale* (It.), vol.11/12, 2024, pp.181-191, [trans. *Criminal Justice: Monthly Journal of Legal Doctrine, Jurisprudence, and Legislation*].

propagation, National Cryptography Centre, information security, blockchain security, data privacy, cybercrime legislation, cryptographic risk.

**Table of Contents.**

1.  **Introduction. - Cybersecurity threats and counter measures.**

During the early decades of the 21st century, significant efforts have been made to address the broad and complex challenges of implementing cyber protection measures regarding both citizens' privacy and national and international security (MOSCA 2012).

The challenges faced in this field, especially in the age of globalisation marked by rapid and incessant scientific progress (BECK 2012), requires dedicated commitment and constant revision to measures designed to repel increasingly sophisticated attacks on computer systems. Such attacks are not only intended to cause damage but can also aim, for instance, to alter production processes, compromise infrastructures for propaganda, disinformation (LISI 2024, 109), or espionage purposes (TETI 2018, 34), or even to radicalize or surreptitiously influence our beliefs and decisions. An opinion has manifested that the greater the knowledge of others' resources and the consistency of the threat, the lesser is the margin for hostile activity. This is particularly concerning given a truly notable and alarming rise in such activities, with repeated attempts to unlawfully obtain data — often for large economic rewards or to blackmail others — especially as it becomes increasingly difficult to trace the actual identities of the offenders.

It must not be underestimated that counter-operations take place within a virtual domain—namely cyberspace—characterised by networks of interconnected IT infrastructures, including hardware, software, data, users, and the logical relationships established among them. In order to implement an adequate security system, this space itself must be protected. It`s therefore crucial to establish the necessary conditions under which it can be safeguarded - particularly through physical, logical, and procedural security measures - to prevent both voluntary and accidental events. These may include the unauthorised acquisition or transmission of data, unlawful modification or destruction, or improper control, damage, or disruption of the normal functioning of networks, IT systems, or their constituent elements.

To summarise, this represents a new arena for competition — also with geopolitical implications — where predominantly (but not exclusively[2]) economic interests operate in virtual space and are used to launch hostile attacks of various origins against public or private entities. These attacks may even involve fundamental human rights and constitutional values (TACCHI 2011, 68).

In light of the aforementioned continuous and unstoppable scientific progress and the recent innovative communication techniques (FADINI 2000), one must ask whether traditional legislation can adapt itself, including by using updated terminology, to regulate conduct in cyberspace comprehensively and coherently, thereby punishing all forms of deviance while respecting the fundamental principles of criminal law. For instance, consider algorithms (which can be classified by their temporal complexity): finite sequences of executable operations designed to solve specific problems, which can be formalised in computer language through a series of instructions (constituting a program), encoding thought into a language comprehensible to a machine that will execute all the instructions defined in that program (CAGLIOTI 2024).

---

[2] The interest generated by the techniques described in the text is also significant in the strategic military sector. In this regard, see (IEZZI 2023). One should also consider the sector concerning geopolitical conflicts in a period like the present, marked by numerous armed conflicts and ransomware attacks primarily aimed at destabilizing the institutions of a given Country.

In these dimensions, the right to data and personal privacy, solemnly enshrined in Article 8 of the European Convention on Human Rights, also comes into play. According to this article, everyone has the right to respect for their private life and various forms of correspondence, meaning that any interference by public authorities must meet well-defined criteria of necessity and proportionality and must pursue specific purposes, to be exercised appropriately (CONTALDO and MULA 2020).

2. **The evolution of cryptology and steganography: technical and legal aspects.**

Among the sectors affected by the phenomenon under investigation, particular attention should be given to the technical and legal aspects of cryptology, that is, 'hidden writings'. This includes the methods of cryptography and cryptanalysis, as well as steganography and steganalysis. These areas are crucial for assessing the limits within which the current legal framework is capable of preventing and punishing cyberattacks, including those originating from cyberspace.

Cryptography is an interdisciplinary field of knowledge that combines logic, mathematics and computer science. It involves the creation, through various methodologies, of 'hidden' or 'secret' writing, so that the message or the embedded data unintelligible to anyone other than authorised individuals possessing the corresponding decryption key. Its objective is to protect communications or information that are not intended to be understood by third parties, thereby preserving their confidentiality.

Conversely, cryptanalysis is a practice used to decode encrypted material without possessing the decryption key, thereby attempting to break encryption algorithms and compromise their security by exploiting their vulnerabilities and applying complex mathematical formulas. Quite often, the results of cryptanalysis reveal weaknesses in the design or implementation of an algorithm, which can reduce the number of keys that need to be tested on the target ciphertext (i.e. encrypted text) for successful decryption[3].

Steganography should be distinguished from cryptography, as it is a specific technique aimed at concealing communication between the two communicating parties. It dates back to the late 1400s and is the work of Johannes Trithemius, who discusses it in his *Ars per occultam scripturam*. This method allows a message to be hidden within a carrier that facilitates its transport without arousing any suspicion[4]. This technique has also been applied in the field of computing (which is hiding information within a file), giving rise to digital steganography. In such cases, electronic communications include steganographic encoding within a transport layer, which could be, for example, a document or an image file[5].

In the digital field, the aforementioned technique therefore consists of the concealment of information within a file through an occultation process that prevents third parties from understanding the actual meaning of the message.

Steganalysis, on the other hand, has the simpler objective of demonstrating the existence of data rather than extracting the hidden information. This practice employs a process that is inverse to the one typically used in steganography.

Over the centuries, starting from the time when Julius Caesar created a cipher to obscure the content of his dispatches delivered to messengers who transported them to the recipients (ROSSANO 2020, 32 et seq), cryptography has undergone significant changes with the advancement of scientific and technological research. From the simplest symmetric encryption system (monoalphabetic), where the cipher, or 'shuffled', alphabet is obtained by replacing the letters of the original alphabet (plain text) with another letter of the alphabet, a different technique was developed. This new method does not rely on a single key for both encryption and decryption but instead uses two distinct keys to make the system less vulnerable and more resistant to cyberattacks. One key is used for encryption and the other for decryption (the so-called asymmetric systems).

---

[3] With specific reference to the economic and financial sector, see V. DIOTALLEVI – N. CINTIOLI (2023).
[4] A. GELPI (2024, 161) provides noteworthy historical insights.
[5] S. ATERNO (2022, 199 et seq.) also examines the issues related to personal data protection, with particular reference to Regulation (EU) 2016/679 of the European Parliament and Council (GDPR), subsequently implemented by various national legal systems and in Italy by Legislative Decree 10 August 2018, no. 101.

In public key cryptography (PKI) or asymmetric cryptography, there are two correlated keys, known as the public key and the private key. While the former can be freely distributed and used for encryption, the latter, associated with the former, must remain confidential and is used for decryption.

Another particularly widespread process related to encryption to determine whether the data in transit has been altered is hashing. This involves transforming data into a fixed-length string, where the data is converted into a series of letters and numbers generated by a hashing algorithm. It plays a crucial role in verifying the integrity of a message.

A further achievement in the field under examination is represented by the advent of quantum cryptography, built around photon sources capable of emitting pairs of *entangled* photons. In these systems, photon A, which travels towards an optical fibre with a polarisation that is perpendicular to that of photon B, which moves in the opposite direction. This entanglement makes the system theoretically resistant to attacks from hackers (OCCHIPINTI 2008).

Unlike what occurs in classical cryptographic systems, as previously discussed, in the quantum cryptography — grounded in Heisenberg's uncertainty principle [6] — security relies on the fact that any attempt of intrusion by an attacker will inevitably disturb the system. This disturbance immediately reveals the presence of the intrusion, thus prompting the immediate replacement of the encryption key being used.

Due to its physical peculiarities, the quantum cryptographic system is capable of detecting unauthorised access or man-in-the-middle attacks, and it is precisely in this capacity that its remarkable innovative strength lies [7].

---

[6] On this subject, see the essay by E. FILORAMO - C. PASQUERO (2006)
[7] Recently, the EuroQCI project was launched to create a secure communication network based on the technique discussed in the text. To enhance the efficiency of quantum systems, we could refer to Peter Shor, who in 1994 developed an algorithm that solves the problem of factoring integers into prime numbers.

Further advancements in the field have arisen from the emergence of post-quantum cryptography (PQC), which no longer relies on the principles of quantum physics but instead on pure mathematics. The new generation of algorithms is designed to operate on traditional or classical computers, which process calculations sequentially and store data using binary bits, each bit representing either a 0 or a 1.

By contrast, quantum computers employ qubits—systems that behave like subatomic particles—to process data within a range of amplitudes applied to both 0 and 1 simultaneously, rather than being limited to two distinct states (0 and 1). The continuous nature of the qubit's state enables the existence of *superposition*, referring to the capacity of a quantum system to exist in multiple states at once, with each state represented by a complex probability amplitude.[8] This unique ability to correlate particles—*entanglement*—enables the creation of complex quantum states that cannot be described by a mere combination of individual qubit states.

It is precisely the high complexity of these algorithms that makes PQC-based security systems extremely difficult to decipher — even for state-of-the-art quantum computers. For this reason, the international scientific community, led by the U.S. National Institute of Standards and Technology (NIST)[9], is actively engaged in research aimed at addressing and neutralizing cybersecurity risks. As part of this effort, PQC systems are being developed to form the foundation of post-quantum networks and communications, and they can also be used on currently available computing systems.

On 13 August 2024, the Agency officially released the first three post-quantum cryptography standards. The algorithm formerly known as CRYSTALS-Kyber, described by NIST as the primary encryption standard, has now been standardised as ML-KEM (*Module Lattice-based Key*

---

[8] The relevance of cryptography in the criminal justice sector was addressed in Resolution 12863/20 of the Council of the European Union, adopted on 24 November 2020. In alignment with the conclusions of the European Council of 1 and 2 October 2020 (EUCO 13/20 the resolution advocates for strengthening the EU's capacity to defend against cyber threats, ensuring a secure communications environment—particularly through quantum cryptography— guaranteeing lawful access to data for judicial and law enforcement purposes. It also calls for a supportive approach to use of electronic evidence, which is essential for conducting effective cybercrime investigations and for adequately protecting victims.

[9] In the United States, these objectives are similarly pursued by agencies such as the National Institute of Standards and Technology (NIST)—a non-regulatory body within the Department of Commerce responsible for promoting innovation, security, and industrial competitiveness—and the Cybersecurity and Infrastructure Security Agency (CISA), which plays a central role in safeguarding the nation's critical infrastructure.

*Encapsulation Mechanism*), under FIPS 203. The digital signature algorithm CRYSTALS-Dilithium has been standardised as ML-DSA (*Module Lattice-Based Digital Signature Algorithm*), under FIPS 204. A third algorithm, SPHINCS+, serves as an alternative digital signature mechanism and has been designated SLH-DSA (*Stateless Hash-Based Digital Signature Algorithm*), under FIPS 205. A fourth algorithm, initially known as FALCON, is currently under review and is expected to be included in the standard suite in the near future. Like the previous two, it is designed for digital signature applications. Accordingly, once the draft FIPS 206 standard is released, the algorithm will be designated FN-DSA, an acronym for *Fast-Fourier Transform over NTRU-Lattice-Based Digital Signature Algorithm*.

Countries at the forefront of quantum technology development - including the United States, Australia, the United Kingdom, and now also China, are all actively collaborating and heavily investing in perfecting current quantum computing systems. Their efforts focus particularly on improving security and developing new, updated systems that prevent unauthorised access aimed at intercepting public or private data and information that should instead remain confidential[10].

3. **Regulatory gaps and *de iure condendo* perspectives.**

In light of the relentless advancement of scientific and technological innovation, it is necessary to question whether, and to what extent, the current legal framework governing the protection of devices intended to perform any function useful to humans — particularly through the partial or full utilisation of information technologies involving processes of data encoding, decoding, storage, and recording (e.g., binary representation via bits) — is adequately equipped to prevent and combat cyber piracy, hacker attacks, or unauthorised access.

---

[10] Recent geopolitical developments, particularly following the last U.S. presidential election, have raised questions about the future of international collaboration. The Trump administration's tariff policies have impacted the emerging quantum computing market, disrupting supply chains and increasing costs—particularly for quantum hardware components. These policies may also restrict talent mobility and access to rare materials, hindering research and development. Moreover, delays in the establishment of quantum-resistant encryption standards, exacerbated by tariff-related disruptions, could leave sensitive systems vulnerable to cyberattacks. Investor concerns and market volatility are likely to be reflected in increased fluctuations in the stock prices of quantum computing firms.

Italian lawmakers initially addressed these issues at the end of the 20th century with Law No. 547 of 23 December 1993[11]. This law, in response to the widespread phenomenon of computer-related damage, introduced Article 635-*bis* into the Italian Penal Code titled `Damage to Information, Data and Computer Programs`. This crime, modelled on the traditional offence type of criminal damage set out in the Article 635 of the Penal Code[12], punishes any conduct aimed at destruction, deterioration, or the total or partial disabling of IT assets, offering enhanced protection for both tangible and intangible assets, such as software. (F. MAZZA 2007, 353 ).

However, the limitations and shortcomings of this legislative reform soon became apparent. (MANTOVANI 2004, 170). Between 2008 and 2009, new measures were adopted to address these gaps and to fully implement the Council of Europe's Convention on Cybercrime (the `Budapest Convention`, signed on 23 November 2001 and in force since 1 July 2004), which also served as a source of inspiration for Framework Decision 2005/222/JHA of the Council of the European Union. Thus, in a hasty response to the imminent end of the legislative term, Law No. 48 of 18 March 2008 was enacted. Despite its significance, the statute suffered from numerous technical and systemic deficiencies — issues that are, regrettably, frequently observed in the drafting of new legislative provisions[13]. It replaced the original text of Article 635-*bis* of the Italian Penal Code and introduced three new types of offence, establishing a four-tiered system for computer-related damage: damage to computer data of private utility, damage to computer data of public utility, damage to computer or telematic systems of private utility, and damage to computer or telematic systems of public utility.

---

[11] On the subject, see C.PECORELLA (ed. 2006, 140 et seq); P. GALDIERI (1997, 101 et seq), who laid the foundations for the scientific study of the forms of crime referred to in the text.
[12] For further observations on this matter, see A. ROSSI VANNINI (1994, 450 et seq)
[13] See V. ITALIA (2020, 23 et seq), where the defects of the laws are listed (quantity, length, obscurity of the lexicon and fragmentation) with the related consequences that make the administrative activity opaque and slow down and complicate the exercise of the jurisdictional function, compromising the principle of equality.

This framework was intended to curtail conduct that threatens critical infrastructures of democratic societies and even inter-state relations.

In reality, rather than needlessly multiplying offence types within the Penal Code — resulting in an inefficient and excessive proliferation of provisions— an effective legal response could have been achieved by reformulating Article 635-*bis* with a more coherent and technically sound language[14], and incorporating in a suitably concise manner the substantive content of the new offence types.

It must also be noted that the drafting of the revised text of the aforementioned article adopts a casuistic rather than a generalising approach. In an attempt to preclude legislative gaps, the punishable conduct is described using no fewer than five verbs (`destroys, damages, erases, alters, or suppresses`), which substantially overlap in meaning. This redundancy renders it difficult to differentiate the precise scope and operational boundaries of each term.

In the same timeframe, Law No. 85 of 30 June 2009 fully implemented the Prüm Treaty, signed on 27 May 2005 between the Kingdom of Belgium, the Kingdom of Spain, the French Republic, the Grand Duchy of Luxembourg, the Kingdom of the Netherlands, and the Republic of Austria. Article 5 of the law established the National DNA Database and the Central Laboratory for the DNA Database. Furthermore, Article 14 further stipulates a prison sentence of one to three years for any public official who discloses or uses data and information in violation of the strict provisions set out in Chapter II of the aforementioned law (L. MAZZA 2021, 1143).

Subsequent legislative developments include Legislative Decree No. 7 of 15 January 2016, Article 2 (paragraph 1, letters m) to p) ), which further modified the damage-related offenses introduced into the Penal Code by the 2008 reform, and Law No. 90 of June 28, 2024, titled `Provisions on Strengthening National Cybersecurity and Cybercrimes`.

---

[14] On the importance of the appropriate use of linguistic signs in the formulation of provisions, see B. MORTARA GRAVELLI (2001, 86 ), which also draws attention to the aspects concerning their arrangement within legal texts.

Legislative Decree No. 7/2016 revised Article 635-*bis*, paragraph 2, of the Penal Code and related Articles from 635-*ter* to 635-*quinquies*, introducing aggravating factors such as the use of violence, threats, or abuse of position as a system operator, thus increasing the severity of penalties.

Further reform came with Law No. 90 of 28 June 2024, which introduced substantive and procedural changes, particularly in the area of cybercrime. This law saw an increase in penalties, a broader definition of specific intent, the introduction of new aggravating circumstances, and restrictions on the application of mitigating factors for crimes committed via digital technology to obtain unjust advantages or unlawfully access or disrupt computer or telematic communications. This reflects an awareness of the paramount role that information systems and telecommunication networks play in the development of human relations and the fulfilment of important institutional goals.[15]

A preliminary review of the complex new legislation suggests that not all objectives, including the closing of protection gaps, have been fully achieved. This is evident in the introduction of Article 635-*quater.1*, which penalises the `unlawful possession, distribution, and installation of equipment, devices, or computer programs intended to damage or disrupt an IT or telematic system`. This provision is overly detailed, with a problematic reliance on the casuistic method. It lists a broad range of prohibited conduct, including vague terms like `illegally` and `abusively`, which complicates the clear delineation of prohibited sets of conduct.

Similarly, the reformulation of Article 635-*quinquies* of the Penal Code retains problematic language, notably because despite the concerns raised under the previous wording, reproduces the hypothesis of conduct involving computer or telematic systems of public interest and the aim of 'seriously hindering their operation'. The concept of `seriousness` remains undefined, leaving it open to the discretion of the interpreter and jurisprudence, potentially infringing the principle of strict

---

[15] The novelty referred to in the text was commented by F. Lombardi ( (LOMBARDI in press)

legality that governs criminal law. The lack of precise criteria for determining what constitutes serious` interference versus lesser conduct opens the door to interpretative ambiguity[16].

Thus, there is a compelling need for a comprehensive review of these types of offence that have accumulated and confusedly layered in a relatively short timeframe, in order to reformulate and consolidate them into a coherent structure. One potential solution would be to establish a dedicated section within Book II of the Penal Code—following Title VIII-*bis* on crimes against cultural property — specifically addressing offences against IT and telematic assets. This new section could integrate (based on the model set out in Article 635 of the Penal Code), the various provisions concerning the mentioned assets damage, data interception and tampering, while streamlining redundant criminal provisions. Separate articles could then address aggravating and mitigating circumstances. This approach would allow for clearer, more precise, and consistent regulation.

The formulation of such provisions must harmoniously integrate technical and legal elements[17], and should be aligned with those already existing in the Penal Code - particularly those addressing unauthorised access and the unlawful dissemination or installation of IT systems (Articles 615-*bis* et seq.) - as well as with the complex and evolving corpus of special legislation in the field. A key example is Legislative Decree No. 196/2003 and its subsequent amendments, which governs personal data protection (and is commonly referred to as the Data Protection Code[18])[19].

To dismantle the illicit trade in unlawfully obtained or manipulated information — often involving corrupt or compromised officials acting on behalf of foreign agents - it is crucial not only to protect the `authentication credentials and access keys`, but also secure critical entry points such as systems gateways and remote servers which are frequently located outside national borders. This necessitates revisiting the principle of territoriality set out in Article 6 of Penal Code (DI MARTINO 2006). Given the increasingly transnational nature of cybercrime, the universalistic tendencies of the

---

[16] On this point see T. PADOVANI (2014, 7 et seq.), and, more recently D. PERRONE (2019).
[17] P. GALDIERI (2021) highlights the difficulties faced by the Italian lawmakers when coining new rules on the matter so that they do not quickly become obsolete.
[18] Codice della privacy [It.].
[19] See B. LOCORATOLO (2024) with bibliographical and jurisprudential references.

Italian legal system — exemplified by the exceptions found in Article 7 (GABRIELI 1936, 22) and the extraterritorial provisions of Article 518 *undevicies* concerning crimes against cultural property—should be leveraged to address cybercrimes committed abroad (F. MAZZA 2024, 189).

In order to implement the scope of cyber security with a harmonised approach at European level for the protection of networks and information systems, Legislative Decree no. 138 of 4 September 2024 was adopted, which implements Directive (EU) 2022/2555 (NIS 2), which imposes a series of obligations on public administrations and private entities. This decree imposes stringent compliance obligations and mandates improvements in the security of IT and network systems to address emerging risks. It also explicitly repeals the earlier Legislative Decree No. 65 of 18 May 2018, which had transposed Directive (EU) 2016/1148 (NIS), now superseded.

## 4. The establishment of the National Cryptography Centre within the National Cybersecurity Agency in Italy: Role and Institutional Objectives.

Article 10 of Law No. 90 of 28 June 2024 redefines the functions of the Italian National Cybersecurity Agency (*Agenzia per la cybersicurezza nazionale*, ACN), originally established by Decree-Law No. 82 of 14 June 2021, converted with amendments into Law No. 109 of 2 August 2021. The Agency was created to safeguard national interests and security in the cybersecurity domain and to facilitate the implementation of the National Recovery and Resilience Plan, approved by the Council of Ministers on 29 April 2021.

The ACN is headed by a Director General and governed by internal governance provisions that define its organisational structure and operational procedures. It coordinates all public bodies involved in cybersecurity and promotes joint activities aimed at enhancing national cyber resilience to support Italy`s digital transformation and the development of its manufacturing and industrial systems, including processes of strategic relevance to the national defence sector[20].

In particular, with regard to cryptography, the newly amended Article assigns the Agency an additional role[21]. By replacing letter m-*bis* of Article 7, paragraph 1, of the aforementioned Decree-Law No. 82/2021, a new section has been established, tasked with assessing the security of cryptographic systems, and with organising and managing awareness raising and outreach initiatives to encourage the adoption of cryptography. These initiatives also highlight the potential benefits of cryptography for blockchain technologies[22], which are increasingly viewed as essential to secure and transparent digital infrastructure.

The same Article 10 formally establishes the National Cryptography Centre, to operate within the CAN. The Centre is to operate without imposing new burdens on public finances—though, in practice, such financial neutrality clauses may prove difficult to uphold, given that further initiatives naturally entail costs [23]). The centre is designated as the national hub for competence in all non-classified cryptographic matters, while the responsibilities of the Central Office for Secrecy (*Ufficio centrale per la segretezza*) remain unaffected.

Given the highly technical and constantly evolving nature of the Agency's mission, it seems appropriate to assign it further competencies—particularly in the prevention and investigation of illicit activities involving cryptographic technologies. This is further justified by the specialised skills of its personnel, for whom the law provides dedicated training and continuing education programs, including partnerships with academic institutions and public or private entities (Article 7, paragraph 1, letter *v*), of Decree-Law No. 82/2021, as amended).

---

[20] Regulation (EU) 460/2004 established with the same purpose the European Network and Information Security Agency (ENISA) with headquarters in Athens, now governed by Regulation (EU) 881/2009 of the European Parliament and of the Council. Recently on 10 October 2024 this Council approved the Cyber Resilience Act (CRA) to improve cybersecurity and cyber resilience.
[21] On the formation of the National Cybersecurity Agency, see S. ATERNO (2022, cit., 234).
[22] Blockchain technology, which originated on 3 January 2009 with the invention of Bitcoin, is characterised by the ability to create shared ledgers incorporating unique properties that offer various opportunities in the business field, providing secure and transparent digital tools for interested parties. On this point, see A. CONTALDO– F. CAMPARA (2019, 37) , as well as M. ROSSANO, (2020, cit.,261).
[23] On the `blocking` function of the financial invariance clause, with reference also to art. 97 of the Constitution, see P. TAVERNITI (2024, cit., 249).

At the international level, the growing importance of digital technologies—particularly in the field of communications—and their rapid scientific advancement have also been acknowledged by the United Nations. In August 2024, the UN approved a draft Comprehensive International Convention[24] on Cybercrime, aimed at establishing a global legal framework to combat cybercrime, enhance international cooperation, and strengthen the capacities of member states to prevent, investigate, and prosecute such offences. (CYBERCRIME (6/2023); (MATTARELLA 2022, 41).

*The Comprehensive International Convention on Countering the Use of Information and Communications Technologies for Criminal Purposes* defines various categories of crimes associated with the use of information and communication technologies (ICT). It also envisions a global monitoring system to periodically update protective measures against emerging threats and attacks — particularly those posed by organised crime — including cyber fraud, data theft, privacy breaches, and malware propagation. At the same time, the Convention emphasises the need to strike a balance between effective crime prevention and combating these crimes, and respecting fundamental human rights and freedoms, such as privacy (MARTORANA 2022) and freedoms of expression (BASSINI 2019).

Ensuring the security of data, emails, web browsing and emerging technologies such as blockchain - a decentralised P2P network governed solely by its own protocol - and cryptocurrencies, which are protected by cryptography to prevent counterfeiting and duplication of the monetary unit ensuring that its holder is unique, is a crucial goal. These digital assets, used for purchases and exchanges through dedicated platforms (CAPACCIOLI 2021), require post-quantum security measures, supported by appropriate legislation, as well as the commitment of all stakeholders—both public and private — engaged in the field of information technology.

---

[24] The Convention will be open for signature at a ceremony to be held in Hanoi, Vietnam, in 2025, and thereafter at United Nations Headquarters in New York until 31 December 2026. The Convention will enter into force after the deposit of ratifications, acceptances, approvals, or accessions by forty states.

## 5. Conclusions

This study has examined the interplay between cryptologic innovation and cybersecurity from a legal-technical perspective, with particular attention to regulatory developments in Italy and European Union. The growing convergence of quantum technologies, cryptographic resilience, and legal oversight highlights both the promise and the vulnerability embedded in modern digital infrastructure.

Recent regulatory advancement in Italy — culminating in Law No. 90/2024, which enhances the cybersecurity safeguards originally introduced in 2008 and establishes the National Cryptography Centre — represent important steps towards institutionalising Italian national cyber defence capabilities. Nonetheless, persistent challenges remain, including casuistic legislative drafting, fragmented criminal provisions, and inconsistencies in regulatory terminology. These shortcomings undermine legal certainty and the principle of strict legality.

To address increasingly complex and transnational cyber threats, it is essential that lawmakers adopt a systematic and coherent penal framework, consolidating existing provisions within a comprehensive structure dedicated to cyber offences. Simultaneously, harmonisation across EU member states—through mechanisms such as the NIS2 Directive— must be pursued to ensure a unified security posture. At the international level, emerging consensus—exemplified by the draft United Nations Convention on Cybercrime—must carefully balance security imperatives and the protection of fundamental human rights, including privacy and freedom of expression.

Forthcoming regulatory reforms should be grounded in technical literacy, interoperable standards, and anticipatory governance - particularly concerning the post-quantum cryptography and blockchain technologies. Only through cross-sectoral coordination and regulatory precision can a secure, resilient, and human rights-respecting digital society be effectively realised.